\documentclass{iaus}
\usepackage{graphicx}

\title[Are the most metal-poor galaxies young?] 
{Are the most metal-poor galaxies young?}

\author[Kunth \& \"Ostlin]   
{Daniel Kunth$^1$%
and G\"oran \"Ostlin$^2$}

\affiliation{$^1$ Institut d'Astrophysique de Paris, 98bis Boulevard Arago, F75014 Paris, France \break email: kunth@iap.fr\\[\affilskip]
$^2$ Stockholm observatory, AlbaNova University Center, 10691 Stockholm, Sweden \break email: ostlin@astro.su.se}

\pubyear{2006}
\volume{235}  
\pagerange{1--4}
\date{20 oct 2006 and in revised form ??}
\jname{Galaxy Evolution across the Hubble Time}
\editors{F.Combes \& J. Palous, eds.}
\begin{document}

\maketitle

\begin{abstract}
 We review the possibility that metallicity  could provide a diagnostic for the age of a galaxy,
 hence that the most metal-poor star forming galaxies in the local universe may be genuinely
 young. Indeed, observational evidence for ``downsizing'' shows the average age of the stars in
 a galaxy to decrease with decreasing mass and metallicity. However,  we conclude both from 
 observational and theoretical viewpoints that metallicity is not an arrow of time. Consequently 
 the  most metal--poor galaxies of our local universe are not necessarely  young. Current 
 observations suggest that an old stellar population is present in all metal-poor galaxies,
 although a couple of cases, e.g. IZw18, remain under debate. Further observations with more 
 sentitive equipement should settle this question in the coming years.
\end{abstract}

\firstsection 

\section{What is a young galaxy?}

In view of the recent popularity of the downsizing scenarios, in which
mean stellar age and metallicity decrease with decreasing galactic mass
 it might be tempting to conjecture that some metal-poor,
  actively star forming dwarf galaxies may be genuinely young, now
forming their first generation of stars. On  the other hand, some of 
the most metal-poor galaxies in our vicinity
are found among the dSph satellites of the Milky Way, which have 
 old stellar populations. Hence, it is quite obvious that 
metallicity cannot be used as a simple cosmic clock, and we can 
straightforwardly answer the title question of this invited talk with a firm No.
   However, the bare possibility that {\it some} local galaxies may
be young calls for a little more scrutiny in answering to this question.

 What criteria must be met before we decide that a clump
of matter is to be classified as a galaxy? Even if galaxies mostly
consists of dark matter, we classify them from their  stellar population 
and  are   inclined towards
using stars and the star formation history as the relevant 
temporal measure for galaxy formation. 
Galaxies do not live in isolation: they merge,
tear out gas and stars from each other, and  accrete  matter. 
What is the proper age definition for a galaxy in 
this context? We take here the working definition that 
the age of a galaxy is equal to that of its oldest stars.

   This subject has been reviewed by \cite{ko2000} and \cite{ko2001}, and
we reiterate one conclusion from this work: finding a genuinely young 
galaxy would be such an interesting discovery that the burden of proof
must lie on the youth hypothesis. Only when it has been shown beyond 
reasonable doubt, the limit being set by current technology, that a galaxy
contains no old stars should we term a galaxy young.

\section{Which are the most metal-poor galaxies?}

Among local actively star forming galaxies, most are dwarfs: dwarf irregulars 
(dIrr) and  Blue Compact Dwarfs (BCD). They remain, however, a minority among 
the general  population of dwarf galaxies, \cite{ko2000}.
At optical wavelengths, their spectra are dominated by young stars and
ionised gas, closely resembling those of giant  HII regions in nearby spiral 
and irregular galaxies. Analysis of their spectra show that most of them 
are metal deficient. Towards the lower end of the metallicity distribution
(O/H $\sim 1/50 Z_\odot$) we find galaxies like IZw18 and SBS0335-052.
Due to their extreme properties it has been conjectured  by \cite{ss1970} 
that these chemically unevolved galaxies could be young systems still in
the proceess of forming. However, dwarf spheroidal galaxies (dSph) can be 
even more  metal-poor, if judged by [Fe/H] for the stellar population, but we 
should keep in mind that this is not directly comparable to nebular [O/H] - 
\cite{gr2003}. Yet, dSphs are undisputably old.

Since metal-poor galaxies exist and are common at low redshift, one could 
of course expect many more at high redshifts. 
The first generation of 
star forming galaxies must, by definition, have been both young  and metal-poor.
However, selection effects tend to discriminate against the smallest, likely 
most metal-poor, galaxies in the distant universe, and the
rather few estimates  for Lyman Break Galaxies
point to rather 'normal' abundances (say 1/3 to $ Z_\odot$) - \cite{pe2006}. 

We note that many parameters control the observed metallicity in  a given galaxy. 
They are incorporated into chemical evolutionnary models and 
include : stellar evolution and nucleosynthesis, inflows and outflows
and the problem of the mixing and dispersion time scale of freshly released 
heavy elements. \cite{ko2000}  caution the use of simple
classical closed box model although clearly, gas mass fraction seems 
to play a major role
\cite{le1979}. 
Moreover metallicity measurements may be relevant to only one 
particular component of a galaxy. A suggestion of  \cite{ku1986}
has been made that HII gas could enrich itself with metals expelled by CC-SNe of time 
scales shorter than 
the  life time of a starburst. Recent FUSE observations of some dwarf galaxies
show a possible disconnect in metallicity between HI and HII regions - \cite{ca2005}, \cite{le2006} - 
although  possible
saturation effects on the line of sight may alter such a comparison - \cite{le2006}.

\section{Metallicity as an arrow of time}
The metal content of galaxies being an important diagnostic of galactic evolution,
the  question  is whether metallicity can be regarded as an arrow 
of time? A useful connection between evolutionary state and metallicity is the 
luminosity--metallicity relationship. 
     Such a relation can naturally arise if smaller galaxies 
have larger gas mass fractions than larger galaxies, i.e. simply because low mass
galaxies have been less efficient in forming stars  and therefore are 
less evolved. 
     Another more likely possibility - see \cite{tr2004} - is that more massive galaxies 
     can retain  metals 
more easily than less massive galaxies because of their deeper gravitational wells, 
making them less susceptible to metal loss in galactic winds - \cite{ga2002}.

 
Numerical simulations by \cite{ce1999} predicting the evolution of the 
metal content of the Universe 
show that metallicity is  a stronger
function of density than age; moreover with a considerable scatter. 
At low redshift, one would expect a few percent of the gas-rich dwarfs to 
have metallicity on the order of IZw18, without invoking their youth.


In the past, the scatter in the N/O  versus O/H
 diagram had been considered to be larger than the observational uncertainties. 
 Time delays between 
the production of oxygen due to massive stars and that of nitrogen
 were a likely part of the explanation although this point of view was
 challenged by  \cite{iz1999}. 
Indeed their observations  not only suggested a small intrinsic 
dispersion of log N/O ($\pm 0.02$ dex) at low metallicities but a similar behaviour 
was found for C/O and some other ratios.  They  concluded that
galaxies with such low abundances are genuinely young (less than 40 Myr old), 
now making their first generation of stars. Moreover they claimed that 
all galaxies with 7.6 $\le$ 12+log(O/H) $\le$ 8.2 have ages from 100 to 500 Myr.
However, independent data such as the CMDs suggest that these galaxies do in fact contain 
old stars and there are definitely many BCDs with  12+ $\log({\rm O/H}) < 8.2$ 
which have  been demonstrated to be much older than 500 Myr 
(see next section). Nevertheless, the scatter is 
surprisingly small  considering the
short time scale for the production  of oxygen (as compared to iron
production) because different stellar masses are involved. 
Now, more recent data using SDSS show a much larger increase of the scatter 
at low (O/H) - \cite{iz2006} while other interpretations for the N/O behaviour are also possible. 
A similar pattern is seen for N/O observations of H{\sc ii} regions in spirals - \cite{ze1998},
and old low surface brightness galaxies - \cite{ro1995}; \cite{be1999}, have N/O comparable
to those of the most metal-poor BCGs.   

\section{Ages of metal-poor galaxies}

Colour-magnitude diagrams (CMDs) allows for the most direct determination
of the age of a stellar population. It is important to point out that in all 
nearby galaxies where CMDs have been obtained to the depth required for a clear
detection of the red giant branch, has this also been found, showing 
beyond doubt that these systems are not young 

Following \cite{iz1999} a galaxy such as IZw18, with its record low 
abundance (12+log(O/H)=7.18), should be genuinely young.
There are nevertheless  independent evidence suggesting that IZw18 does
indeed host an old underlying population (see M. Tosi, this volume).
A careful analysis of archival HST/WFPC2 data  (\cite{al1999}), as well
as independent HST/NICMOS observations  (\cite{ostlin2000}, \cite{om2005}), 
both give support for an age in excess of 0.5  Gyr. 
    Using new, very sensitive, HST/ACS observations \cite{iz2004} failed 
 to trace an old RGB in IZw18, but a re-analysis of 
the same data by \cite{mo2005} of IZw18 do suggest the presence of
an older RGB with an age of several Gyrs.

In cases where galaxies are too distant for a CMD to be obtained,
other ways have to be sought to estimate their age. 
One such method involves using star clusters -  
usually  good representations of single stellar populations 
whose ages can be determined with less degeneracies than
 mixed stellar populations. Limitations are rather set by statistical effects 
 (in dwarf galaxies suitable massive
star clusters are rare). One example of a metal-poor galaxy whose
age can be tied to $\sim 10$ Gyr is the luminous compact galaxy
ESO\,338--04, a.k.a. Tol\,1924-416 (\"Ostlin et al. 1998).

As  most active star formation in metal-poor galaxies tends to
be centrally concentrated, a natural place to look for old stars
are the outskirts, or ``haloes'' of e.g. BCDs  by comparing 
optical and infrared colours to spectral synthesis models
(e.g. Bergvall \& \"Ostlin 2002). The further out
one looks, the smaller the contamination from young stars, but also
 lesser is the light  to analyse. A major problem is that of
wide spread ionised gas.
 In addition, the
ionised gas may have a different scale length from the stellar
population (Papaderos et al. 2002), further complicating the analysis.

With good quality data it is even
possible to date the old population in the centra of star
forming galaxies, e.g. as has been performed with some SDSS spectra
of star forming dwarfs (Corbin et al. 2006).

\section{Outlook}

Our discussion lead us to conclude that no compelling observational evidence
for any young local galaxy has been provided so far. Moreover metallicity is 
not an arrow of time, but a product of the past star formation history, combined
with the mass-loss and accretion history. Only few local galaxies 
remain under debate but this is likely because 
they remain beyond sensitivity thresholds!
With JWST we can probe a little further, but not much, e.g. getting a CMD 
for SBS\,0335-052 will still be a tough task.
Do we at all expect young galaxies in our Local Universe? 
Could some  HI  clouds that  survived reionisation start to collapse now? 
Then they need to have been unable to form any stars early on but still
be masssive enough not to boil off completely during 
reionisation. It appears unlikely that both requirements can be fulfilled.

\begin{acknowledgments}
We wrote this paper for money.
\end{acknowledgments}


\begin{thebibliography}{}


\bibitem[Aloisi et al.(1999)]{al1999} 
				{Aloisi, A., Tosi, M., \& Greggio, L.}, 1999,
				\textit{AJ}, 118, 302

\bibitem[Bergvall \& \"Ostlin (2002)]{be2002} 
         {Bergvall, N.,  \"Ostlin, G.}, 1999, 
         \textit{A\&A}, 390, 891

\bibitem[Bergvall et al. (1999)]{be1999} 
         {Bergvall, N., R"onnback, J., Masegosa, J., \"Ostlin, G.}, 1999, 
         \textit{A\&A}, 341, 697 
         
         
\bibitem[Cannon et al. (2005)]{ca2005} 
        {Cannon, J.~M., Skillman, E.~D., Sembach, K.~R.,
         \& Bomans, D.~J.}, 2005,
        \textit{ApJ}, 618, 247 

\bibitem[Cen \& Ostriker (1999)]{ce1999}
      {Cen R., Ostriker J.P.}, 1999,
      \textit{ApJ}, 519, L109
        
\bibitem[Corbin et al. (2006)]{corbin}
         {Corbin M., Vacca W., Cid Fernandes R., et al.}, 2006, 
         \textit{ApJ},  in press.

\bibitem[Dufour et al.(1996)]{du1996} 
      {Dufour, R.~J., Garnett, D.~R., Skillman, E.~D., \& Shields, G.~A.}, 1996, 
       \textit{Science with the Hubble Space Telescope} - II, 348
    
			
\bibitem[Garnett (2002)]{ga2002}
      {Garnett, D.~R.}, 2002, 
      \textit{ApJ}, 581, 1019        
       
\bibitem[Grebel et al. (2003)]{gr2003} 
      {Grebel, E.~K., Gallagher, J.~S., III, \& Harbeck, D.}, 2003,
      \textit{AJ}, 125, 1926 

\bibitem[Hunter \& Thronson(1995)]{hu1995} 
			{Hunter, D.~A., \& Thronson, H.~A., Jr.}, 1995, 
			\textit{ApJ},  452, 238 
			
\bibitem[Izotov \& Thuan (1999)]{iz1999} 
      {Izotov, Y.~I., \& Thuan, T.~X.}, 1999, 
      \textit{ApJ}, 511, 639 
      
\bibitem[Izotov \& Thuan(2004)]{iz2004} 
      {Izotov, Y.~I., \& Thuan, T.~X.}, 2004, 
      \textit{ApJ}, 616, 768 
 
\bibitem[Izotov et al.(2006)]{iz2006} 
			{Izotov, Y.~I., Stasi{\'n}ska, G., Meynet, G., Guseva, N.~G., \& Thuan, T.~X.}, 2006, 
			\textit{ApJ}, 448, 955 
			
        
\bibitem[Kunth \& \"Ostlin (2000)]{ko2000}
     {Kunth D.,  \"Ostlin G.}, 2000,  
     \textit{A\&ARv}, 10, 1

\bibitem[Kunth \& \"Ostlin (2001)]{ko2001}
     {Kunth D.,  \"Ostlin G.}, 2001,  
     \textit{ApSSS }, 277, 281
     
\bibitem[Kunth \& Sargent (1986)]{ku1986} 
     {Kunth, D., \& Sargent, W.~L.~W.}, 1986, 
     \textit{ApJ}, 300, 496 
     
\bibitem[Lebouteiller et al.(2006)]{le2006} 
     {Lebouteiller, V., Kunth, D., Lequeux, J., Aloisi, A., Desert, 
     J.~-., Hebrard, G., Lecavelier des Etangs, A., \& Vidal-Madjar, A.}, 2006,
     \textit{Astrophysics e-prints}, astro-ph/0608445     
     
\bibitem[Lequeux et al. (1981)]{le1981} 
      {Lequeux, J., Maucherat-Joubert, M., Deharveng, J.~M., \& Kunth, D.}, 1981,
      \textit{A\&A}, 103, 305 
      
\bibitem[Lequeux et al. (1979)]{le1979} 
      {Lequeux, J., Peimbert, M.; Rayo, J. F.; Serrano, A.; Torres-Peimbert} 1979,
      \textit{A\&A}, 80, 155 
      
\bibitem[Momany et al.(2005)]{mo2005} 
       {Momany, Y., et al.}, 2005,
       \textit{ApJ},  439, 111 
       
\bibitem[\"Ostlin et al. (1998)]{os1998} 
     {\"Ostlin, G., Bergvall, N., \& Roennback, J.}, 1998, 
     \textit{A\&A}, 335, 85 
     
\bibitem[\"Ostlin (2000)]{ostlin2000}
     { \"Ostlin G.}, 2000,  
     \textit{ApJ}, 535, L99

\bibitem[\"Ostlin \& Mouhcine (2005)]{om2005}
     {\"Ostlin G., Mouhcine, M.}, 2005,  
     \textit{A\&A}, 433, 979

\bibitem[Papaderos et al. (2002)]{papa} 
      {Papaderos P., Izotov Y., Thuan T., et al.}, 2002,
      \textit{A\&A}, 393, 461

\bibitem[Pettini(2006)]{pe2006} 
      {Pettini, M.}, 2006, 
      \textit{ASP Conf.~Ser.}, 353, 363 
      
\bibitem[R\"onnback \& Bergvall (1995)]{ro1995} 
     {R\"onnback, J., \& Bergvall, N.}, 1995, 
      \textit{A\&A}, 302, 353 
      
\bibitem[Sargent \& Searle (1970)]{ss1970}
        {Sargent, W.~L.~W., \& Searle, L.}, 1970,
        \textit {ApJ}, 162, L155 
        
\bibitem[Tremonti et al. (2004)]{tr2004} 
         {Tremonti, C.~A., et al.}, 2004, 
         \textit{ApJ}, 613, 898
                 
\bibitem[van Zee et al.(1998)]{ze1998} 
	{van Zee, L., Salzer, J.~J., \& Haynes, M.~P.}, 1998, 
				\textit {ApJ}, 497, L1 
	
			                
\end{thebibliography}
\end{document}